\documentclass{iau}
\usepackage{graphicx}
\usepackage{natbib}

\title[Projection Effects in TNG Shell Galaxies] %% give here short title %%
{Projection Effects in Merger Dating \\ for Illustris TNG Shell Galaxies }

\author[Erol, T., Yıldız, M.K., Ebrová, I., Bílek,M.]   %% give here short author list %%
{Tuğba EROL$^{1,2}$
, Mustafa Kürşad YILDIZ$^{2,3}$, Ivana Ebrová$^3$, Michal Bílek$^4$}

\affiliation{$^1$Erciyes University, Graduate School of Natural and Applied Sciences, Kayseri, Türkiye \\ email: {\tt terol@erciyes.edu.tr} \\[\affilskip]
$^2$Erciyes University, Faculty of Science, Astronomy and Space Science Department, Kayseri, Türkiye \\%email: {\tt mkyildiz@erciyes.edu.tr} \\
$^3$Erciyes University, Astronomy and Space Sciences Observatory Applied and Research Center (UZAYBİMER), Kayseri, Türkiye \\
$^4$FZU – Institute of Physics of the Czech Academy of Sciences, Na Slovance 1999/2, Prague 182 21, Czech Republic \\ %email: {\tt ebrova.ivana@gmail.com} \\[\affilskip]
$^5$Astronomical Observatory Belgrade, Volgina 7, 11060 Belgrade, Serbia %\\ email: {\tt michal.bilek@aob.rs}

}

\pubyear{2026}
\volume{403}  %% insert here IAU Symposium No.
\pagerange{119--126}
\jname{The Hidden Beauty of the Galactic Outskirts}
\editors{}
\begin{document}

\maketitle

\begin{abstract}
Stellar shells are low-surface-brightness structures that typically appear as concentric arcs. They are observed in many giant elliptical and lenticular galaxies, as well as in some spiral and dwarf galaxies. 
They presumably result from minor and intermediate close-to-radial mergers of galaxies. 
An essential factor in determining the merger time are the distances of shells from the centre of the galaxy. 
Consequently, estimates of the merger time can be significantly affected if some shells remain undetected due to projection effects. 
In this study, we present the first systematic investigation of how measured shell radii depend on the orientation of a galaxy relative to the observer.
Using the Illustris TNG50 simulation, we examine shell galaxies from nine selected lines of sight and measure the shell radii. 
We model shell evolution and calculate the merger time accordingly for each viewing angle and quantify the impact of orientation on the inferred merger age. 
Our results indicate that the line of sight can have a substantial effect on merger-time estimates derived from shell radii.

\end{abstract}

\firstsection % if your document starts with a section,
\section{Introduction}

Galaxy mergers are one of the most important interactions determining galaxy evolution \citep[][and references therein]{2012A&A...539A..45L}.
These interactions occur through the merger of two or more galaxies due to gravity.
During these mergers, various structures such as 'tails', 'streams', 'asymmetric halos', 'double nuclei', and 'shells' are formed due to the gravitational tidal effects that galaxies exert on each other.
There are many studies that aimed to understand shell and their relation with galaxy mergers and evolution \citep[e.g.,][]{malin,sw,weil,sikkema,Duc_2014,B_lek_2016}. 

Shells are concentric arc-like stellar structures around galaxies \citep{arp66}.
These structures do not intersect each other and are composed of stars originating from the less massive progenitor, which are deposited on nearly radial orbits \citep{Quinn_1983}. In such a scenario, first shells emerge close to the host galaxy center and they continuously expand away from the host galaxy center.
After a certain time, the shell structure disperses and disappears \citep{prieur}. 

Shells in shell galaxies are numbered from the outside inward, as the first shell to form is also the outermost one. Consequently, the galactocentric radius of the outermost shell is the key observable for estimating the time since the merger, providing a lower limit on the merger age \citep{Ebrov__2020}.
We can observe galaxies only from a single line of sight. 
However, the detectability of a shell strongly depends on the viewing angle \citep{2015CaJPh..93..203B}. 
Some shells may therefore be obscured or remain undetected, leading to biased estimates of shell radii and merger times. The magnitude of this uncertainty has not yet been systematically investigated or quantified.

Different lines of sight can be accessed using cosmological simulations. 
Rather than simulating individual shell galaxies from idealized initial conditions, we select shell galaxies that have formed naturally in a cosmological simulation and examine them from a wide range of lines of sight. This approach allows us to investigate how the apparent distribution and detectability of shells depend on viewing angle, how the number of visible shells changes with orientation, and ultimately how these effects influence estimates of merger times.
In this study, we investigated how the distribution of shells depends on the line of sight and how the number of visible shells changes with viewing angle.

\begin{figure}[ht!]
% \vspace*{-2.0 cm}
\begin{center}
 \includegraphics[width=5.3in]{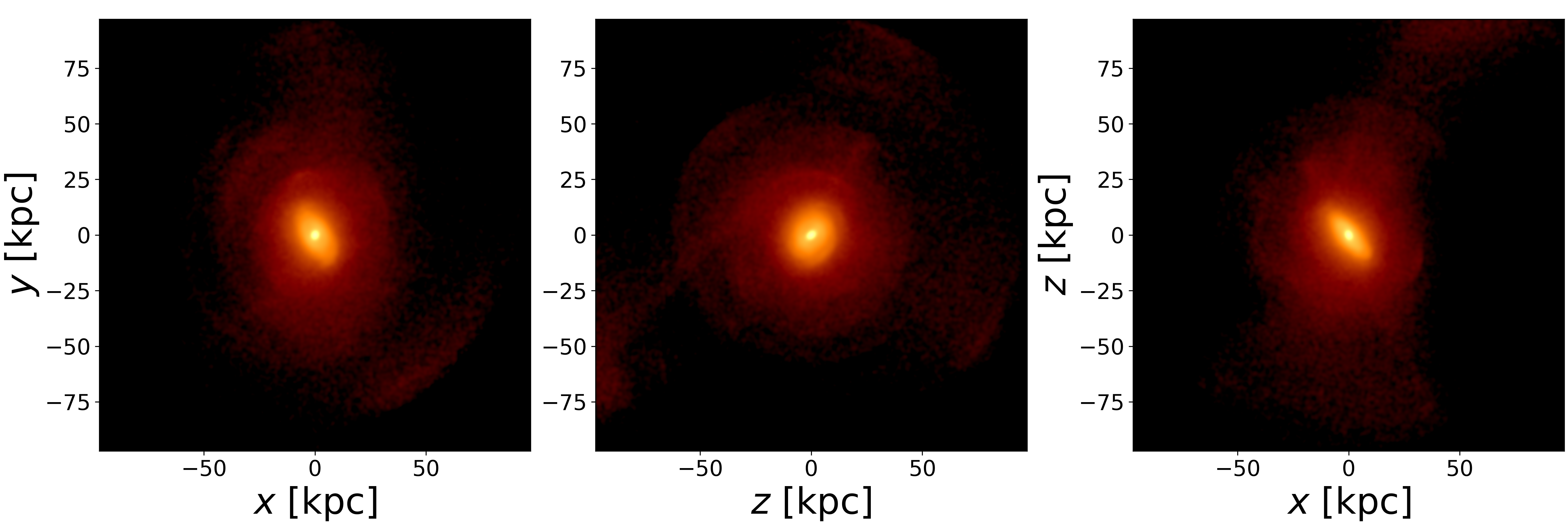} 
% \vspace*{-1.0 cm}
 \caption{Surface brightness maps of three orthogonal projections of TNG50 SubID=511920; the first panel corresponds to the viewing angle ‘0’ as labeld in Table\,\ref{tab:combined_label}.}
   \label{fig:tng50}
\end{center}
 \vspace*{-0.5 cm}
\end{figure}

\section{Method}
\label{sec:Yöntem}

In this study, we use the TNG50 simulation \citep{Nelson_2017,Springel_2017,Naiman_2018,Pillepich_2017,Marinacci_2018}, which combines sufficiently high spatial resolution to identify and examine stellar shells with a simulated volume spanning approximately 50 Mpc on a side \citep{Nelson_2019,Pillepich_2019}. 
The simulation stores the properties of galaxies in a series of snapshots at different cosmic times. Since galaxies are three-dimensional objects, they can be viewed from different directions. For each selected shell galaxy, we therefore create projected images along multiple lines of sight. 
We use nine different points of view that cover This allows us to investigate how the appearance, detectability, and measured radii of shells depend on the viewing angle.

We use nine different points of view that provide a representative sampling of viewing angles. Since galaxies are effectively optically thin in this context, opposite directions yield nearly identical projections. Therefore, this set of viewpoints captures the relevant range of distinct orientations needed to study the dependence of shell visibility on the line of sight.
The angles are listed in Table\,\ref{tab:combined_label}.

\section{Results and Discussion}
\label{sec:3}

\begin{figure}[]
% \vspace*{-2.0 cm}
\begin{center}
 \includegraphics[width=5.4in]{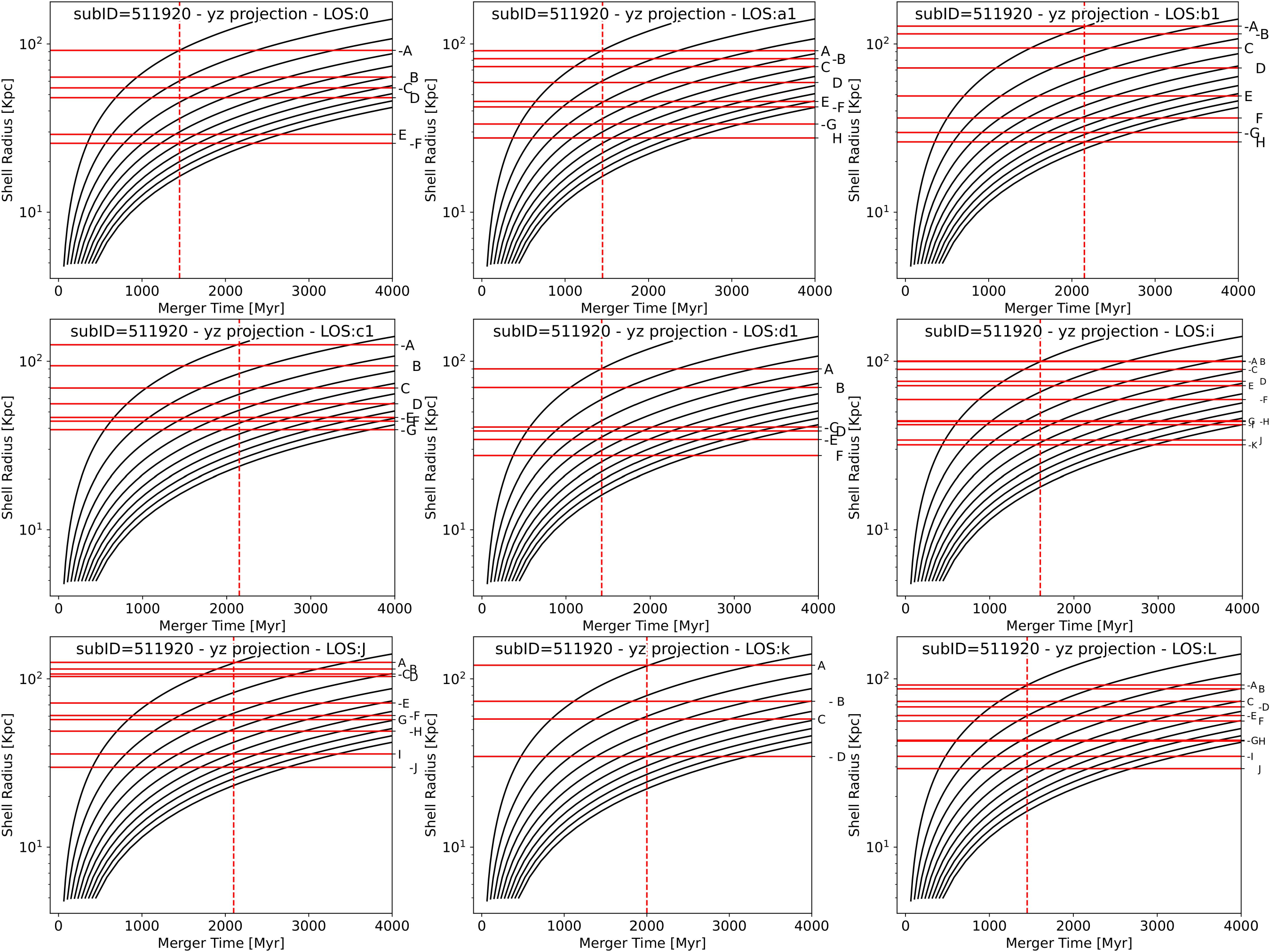} 
% \vspace*{-1.0 cm}
 \caption{Black curves show the modeled evolution of the shell radii within the gravitational potential of a selected TNG50 galaxy. Red lines correspond to the shell radii detected in mock images in nine different projections. 
The dashed vertical lines mark lower merger time estimates for each projection; see also Table\,\ref{tab:combined_label}. 
 }
   \label{fig:mesh9}
\end{center}
\end{figure}

Here we present an example of a TNG50 shell galaxy, see Figure\,\ref{fig:tng50}. 
We use our mock images to detect the maximum of the visible shells in each of the nine projections. 
Measured shell distributions are visualized as red horizontal lines in Figure\,\ref{fig:mesh9}. 
Black curves show modelled evolution of shell radii in the gravitational potential of the selected galaxy \cite[see e.g.,][]{Ebrov__2020, Bilek_474}. 
Comparison of the measured and modelled shell distribution can be used to more accurate merger time estimates by employing the ‘shell identification method’ \citep{Bilek_2013,Bilek_2014}. 
However, in this study, we focus on the robust estimates of the lower limit of the merger time. 
This estimate is based on comparing the outermost modelled and measured shell. 
The correction of those two gives the absolute minimum time that has elapsed since the merger.

The inferred lower time estimates for each projection are depicted in Figure\,\ref{fig:mesh9} as red dashed vertical lines and listed in Table\,\ref{tab:combined_label}. 
Although the outermost shell still gives a robust lower limit of the merger time, for a single galaxy this estimate can vary by up to 0.7\,Gyr depending on the line of sight. 
This level of variation is expected based on previous studies, which have demonstrated that the apparent positions of shells depend on the viewing angle. However, the magnitude of this effect has not previously been quantified using simulations. Since observational studies are inherently limited to a single line of sight, this source of uncertainty must be taken into account when interpreting merger time estimates. 

We also see that the distribution of shell radii itself changes significantly in different projections. This can affect attempts for the full shell identifications. In our on-going study (Erol et al., in prep.), we investigate the impact of the projection effects on such merger time estimates for TNG50 shell galaxies.

\begin{table}[t]
    \centering
    \begin{tabular}{|c|c|c|c|} 
        \hline
        \textbf{Line of Sight} & \textbf{Theta ($\theta$)} & \textbf{Phi ($\phi$)} & \textbf{Merger Time Lower Limit} \\ \hline 
        \textbf{0}  & $\pi$*0=0 & $\pi$*0=0 & 1.45 Gyr \\ 
        \textbf{a1} & $\pi$/8=$22.5^\circ$ & $\pi$/4=$45^\circ$ & 1.45 Gyr \\ 
        \textbf{b1} & 5*$\pi$/8=$112.5^\circ$ & $\pi$/4=$45^\circ$ & 2.15 Gyr \\ 
        \textbf{c1} & 9*$\pi$/8=$202.5^\circ$ & $\pi$/4=$45^\circ$ & 2.15 Gyr \\ 
        \textbf{d1} & 13*$\pi$/8=$292.5^\circ$ & $\pi$/4=$45^\circ$ & 1.42 Gyr \\ 
        \textbf{i}  & $\pi$*0=0 & $\pi$/2=$90^\circ$ & 1.6 Gyr \\ 
        \textbf{j}  & $\pi$/4=$45^\circ$ & $\pi$/2=$90^\circ$ & 2.1 Gyr \\ 
        \textbf{k}  & $\pi$/2=$90^\circ$ & $\pi$/2=$90^\circ$ & 2.0 Gyr \\ 
        \textbf{l}  & 3$\pi$/4=$135^\circ$ & $\pi$/2=$90^\circ$ & 1.45 Gyr \\ \hline
    \end{tabular}
    \caption{Viewing angles and the lower limits of merger times for the nine given lines of sight.}
    \label{tab:combined_label}
\end{table}

\section*{Acknowledgements}
We acknowledges the support from the Postgraduate and Specialist Thesis Project, supported by the Scientific Research Projects Unit at Erciyes University / Türkiye (No. FYL-2024-13884) (TE); the FORTE project, supported by the Ministry of Education, Youth, and Sports of the Czech Republic, co-funded by the European Union (No. CZ.02.01.01/00/22\_008/0004632) (IE); and the Ministry of Science, Technological Development and Innovation of the Republic of Serbia under contract no. 451-03-33/2026-03/200002 with the Astronomical Observatory of Belgrade (MB).

\end{document}